\documentclass[pre,aps,amssymb,amsmath,superscriptaddress]{revtex4}
\usepackage{amsmath}
\usepackage{amssymb}
\usepackage{epsfig}
\usepackage{amscd}
\usepackage{graphicx,psfrag,xspace}
\usepackage{float}
\usepackage[normalem]{ulem}

\usepackage{graphicx}

\begin{document}

\title{Pre-asymptotic corrections to fractional diffusion equations}
\author{Marzio Marseguerra}
\author{Andrea Zoia}
\email{andrea.zoia@polimi.it}
\affiliation{Polytechnic of Milan, Nuclear Engineering Department, Milan, 20133 Italy}

\date{\today}

\begin{abstract}
The motion of contaminant particles through complex environments such as fractured rocks or porous sediments is often characterized by anomalous diffusion: the spread of the transported quantity is found to grow sublinearly in time due to the presence of obstacles which hinder particle migration. The asymptotic behavior of these systems is usually well described by fractional diffusion, which provides an elegant and unified framework for modeling anomalous transport. We show that pre-asymptotic corrections to fractional diffusion might become relevant, depending on the microscopic dynamics of the particles. To incorporate these effects, we derive a modified transport equation and validate its effectiveness by a Monte Carlo simulation.
\end{abstract}

\maketitle

\section{Introduction}

A diffusion process is called {\em anomalous} when the variance of the transported quantity, $\left\langle x^2(t) \right\rangle$, grows in time as $t^\alpha$, with $\alpha$ larger (superdiffusion) or smaller (subdiffusion) than one. A general approach to the analysis of stochastic transport is based on the continuous time random walk (CTRW), a probabilistic model where the motion of a walker in a medium is interpreted as a series of jumps of random lengths, separated by random waiting times \cite{beyondbm, klafter1, klafter2, weiss}. The theory of CTRW with power-law probability distribution functions (pdf's) for the waiting times was first introduced in physics in the late 1960s to describe anomalous diffusion of charge carriers in amorphous semiconductors \cite{scher_lax1, scher_lax2, scher, montroll}. It was later applied with success to a broad spectrum of physical systems (see, for example, Refs.~\cite{klafter1} and~\cite{klafter2} for a review). 

As a paradigmatic example of anomalous diffusion we consider here the case of contaminant transport through heterogeneous materials, such as porous sediments or fractured rocks. In this system the microscopic dynamics of each particle is assumed to be hindered by obstacles, dead ends, trapping events due to the interaction with the surrounding environment or abrupt changes in the flow field. The macroscopic effect is that the spread of the migrating particles plume grows sublinearly in time, thus resulting in {\em subdiffusion} \cite{cortis_gallo, cortis_berkowitz,berkowitz_klafter, berkowitz_kosakowski, berkowitz_scher, berkowitz_scher2, dentz_cortis, levy_berkowitz, margolin}. These assumptions are substantially corroborated by many experimental observations \cite{kimmich, klemm1, berkowitz_klafter, berkowitz_kosakowski, cortis_berkowitz}.

In general, the CTRW equations do not allow for closed-form analytical solutions. However, if suitable first-order approximations are introduced, a fractional diffusion equation can be formally derived from CTRW and analytical solutions can be obtained: in this respect, the fractional diffusion equation represents an asymptotic subset of CTRW \cite{beyondbm, klafter1, klafter2, barkai, schneider, sokolov}.

A complementary approach to the modeling of stochastic transport is provided by Monte Carlo simulation, a particle tracking method which allows the fate of each walker to be followed by sampling the jump lengths and the waiting times from a suitable distribution. Monte Carlo simulation has been widely adopted as a natural tool to obtain reliable and accurate estimates of the CTRW solutions \cite{dentz_cortis}.

In this paper, we explore the range of validity of the asymptotic fractional diffusion equation. This topic has been previously examined e.g. in Refs.~\cite{barkai1} and~\cite{barkai2}. With the help of Monte Carlo simulations, building on the discussion in Refs.~\cite{margolin} and~\cite{marseguerra_zoia1}, we show that pre-asymptotic corrections to the fractional diffusion equations play a significant role, depending on the microscopic dynamics of the particles \footnote{Similar pre-asymptotic corrections have also been proposed for the case of superdiffusion as modelled with L\'evy Flights \cite{zoia_rosso}}. Neglect of these corrections would lead to gross errors in the estimation of model parameters from measured data. To overcome this problem, we derive a modified equation which incorporates these effects.

The paper is organized as follows. In Sec.~\ref{CTRW} we review the CTRW model and the asymptotic fractional diffusion equation. In Sec.~\ref{corrections} we show that for a given range of a critical parameter the fractional diffusion equation might not properly characterize the actual spread of the transported quantity as derived from CTRW (through a Monte Carlo estimate). We discuss how to modify the fractional diffusion model so to include the contribution of pre-asymptotic corrections and validate the proposed results by comparing them to Monte Carlo simulations. Conclusions are discussed in Sec.~\ref{conclusions}.

\section{From CTRW to fractional diffusion}
\label{CTRW}

We briefly review here the main assumptions of the CTRW model. Consider a walker whose stochastic trajectory in a medium is modeled as a series of random jumps separated by random waiting times, during which the walker stays in the previously reached position. The associated pdf $P(x,t)$ represents the probability density of the walker being at $x$ at time $t$ and is usually called the {\em propagator} of the process. For contaminant transport, $P(x,t)$ represents the normalized particle concentration. The properties of the propagator depend on the jump lengths pdf $\lambda(x)$ and the waiting times pdf $w(t)$. Once these distributions are assigned, it is possible to derive a probability equation for $P(x,t)$, the {\em master equation} of the CTRW model \cite{klafter1, weiss, scher, montroll} It can be shown \cite{klafter1} that the master equation in $\lbrace x,t\rbrace$ space can be recast into a simple algebraic relation for the Fourier and Laplace transformed propagator $P(k,u)={\cal L}\left\lbrace {\cal F}\lbrace P(x,t)\rbrace \right\rbrace$ \footnote{We adopt the convention of denoting the $\cal F$ and $\cal L$ transform of a function by its argument: the variable $x$ is mapped to $k$ by the Fourier transform and the variable $t$ is mapped to $u$ by the Laplace transform, so that $f(x,t) \to f(k,u)$.}
\begin{equation}
P(k,u)=\frac{1-w(u)}{u}\frac{1}{1-\lambda(k)w(u)},
\label{ctrw_pdf}
\end{equation}
with the walkers starting at $x=0$ at time $t=0$.

Within the CTRW scheme the long waiting times which characterize the trajectory of a contaminant particle moving through heterogeneous materials are taken into account by assuming that $w(t)$ has a power-law tail $w(t) \sim t^{-1-\alpha}$, $\alpha<1$, when $t \rightarrow +\infty$. Therefore, $w(t)$ lacks a characteristic time scale (the average of the distribution is infinite) and anomalously long waiting times have a non-negligible probability of being sampled. In the following we will assume a Pareto pdf of the form
\begin{equation}
w(t)=\alpha \tau^\alpha t^{-1-\alpha}, \qquad (t \ge \tau)
\label{w_t}
\end{equation}
where $\tau$ is a time constant. This specific form for $w(t)$ has been chosen so that Monte Carlo sampling by inverse transform is straightforward: it can be shown that the short time behavior of the pdf is of minor importance and only the tail matters in determining the properties of $w(t)$ \cite{feller}. The jump length pdf $\lambda(x)$ is usually assumed to be a Gaussian distribution, so that the jumps have a typical scale (say the standard deviation of the distribution) and extreme events, that is, jumps whose length are much larger than the standard deviation, are very improbable.

In general, no analytical solution is known for Eq.~(\ref{ctrw_pdf}), because the required inverse transforms are usually not trivial. However, exact results can be obtained for $P(x,t)$ sufficiently far from the origin, in the {\em diffusion limit} $|x| \to +\infty, t \to +\infty$, that is, $k \to 0, u \to 0$ (see, for example, Ref.~\cite{barkai} for a detailed discussion).

If $w(u)={\cal L} \left\lbrace w(t)\right\rbrace$ is expanded in the long time limit $u \tau \ll 1$ and truncated to the first non-constant term in $u$, we find $w(u) \simeq 1- c_\alpha (u \tau)^\alpha$, where $c_\alpha=\Gamma(1-\alpha)$. More generally, it can be shown that any pdf of the kind $w(t) \sim t^{-1-\alpha}$ would lead to the same expansion $w(u)$ in the transformed space; the specific value of the constant $c_\alpha$ depends on the details of the function \cite{feller, klafter1}. In contrast, if $\lambda(x)$ is a Gaussian with mean $\mu = 0$ and variance $2 \sigma^2$, then $\lambda(k)={\cal F}\left\lbrace \lambda(x)\right\rbrace $ becomes $\lambda(k) \simeq 1-k^2\sigma^2$ in the diffusion limit $k \sigma \ll 1$. Any jump pdf with finite variance would asymptotically lead to the same result \cite{klafter1}.

We now introduce the {\em Riemann-Liouville} fractional derivative operator \cite{podlubny, miller_ross, kilbas} $\partial^{1-\alpha}_{0, t}$, which is defined by its action on a sufficiently well behaved function $f$:
\begin{equation}
\partial^{1-\alpha}_{0, t} f(t)=\frac{d}{dt}\partial^{-\alpha}_{0, t}=\frac{1}{\Gamma(\alpha)} \frac{d}{dt} \int_0^t \frac{f(t')}{(t-t')^{1-\alpha}} dt'.
\label{rl}
\end{equation}
The Riemann-Liouville operator is an integro-differential operator involving a convolution integral with a power-law kernel. In particular, the operator $\partial^{-\alpha}_{0, t}$ may be thought as a generalization of the usual multiple integral of integer order. From its definition, $\partial^{-\alpha}_{0, t}$ has a very simple representation in $\cal L$-space: ${\cal L}\left\lbrace \partial^{-\alpha}_{0, t}f(t)\right\rbrace =u^{-\alpha}f(u)$, $\forall \alpha>0$ \cite{podlubny, miller_ross, kilbas}.

If we substitute $w(u)$ and $\lambda(k)$ in Eq.~(\ref{ctrw_pdf}) and make use of the definition (\ref{rl}), we can formally write the propagator (\ref{ctrw_pdf}) in the $\left\lbrace x,t \right\rbrace$ space:
\begin{equation}
\frac{\partial}{\partial t}P(x,t) =D_\alpha \partial^{1-\alpha}_{0, t} \frac{\partial^2}{\partial x^2}P(x,t),
\label{fde_definition}
\end{equation}
where $D_\alpha=\sigma^2/c_\alpha \tau^\alpha$ may be thought as a generalized diffusion coefficient \cite{klafter1}. Equation (\ref{fde_definition}), which is called the {\em fractional diffusion equation} due to its fractional derivative, is a generalization of the classical Fickian diffusion equation and describes the asymptotic behavior of a plume of subdiffusive particles in the absence of advection. A more general fractional differential equation was derived from CTRW in Refs.~\cite{afanasiev} and \cite{zavslasky_chaos}. The power-law kernel in $\partial^{1-\alpha}_{0, t}$ decays slowly and macroscopically represents the long-range correlations induced by the obstacles that the particles encounter \cite{zavslasky_chaos, mainardi}. If $\alpha>1$, the correlations decay sufficiently quickly so that normal diffusion is recovered \cite{klafter1}.

\begin{figure}[t]
\begin{center}
\scalebox{0.60}{\includegraphics{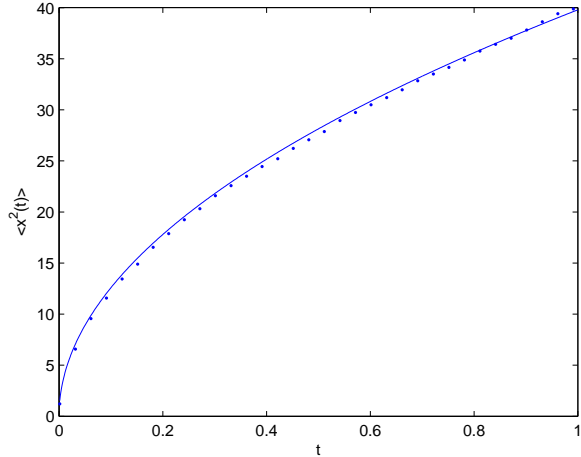}}
\caption{Comparison between the  analytical result (Eq.~(\ref{variance_fde_sub}), solid line) and the Monte Carlo variance (dots) as a function of time for $\alpha=0.5$. The agreement is remarkable.}
\label{fig_fde_sub1}
\end{center}
\end{figure}

In the context of contaminant transport it is of utmost importance to determine the evolution of the plume spread as a function of time. This physical quantity corresponds to the variance of the propagator. From the fractional diffusion equation it may be expressed in closed form as:
\begin{equation}
\left\langle x^2(t) \right\rangle \equiv-\lim_{k \rightarrow 0} \frac{\partial^2}{\partial k^2} {\cal L}^{-1}\left\lbrace P(k,u)\right\rbrace= \frac{2 D_\alpha}{\Gamma(1+\alpha)}t^\alpha.
\label{variance_fde_sub}
\end{equation}
Because $\alpha<1$, the plume migration is subdiffusive, as expected.

\section{Higher-order corrections to fractional diffusion}
\label{corrections}

In principle, Eq.~(\ref{variance_fde_sub}) provides a reliable estimate of the actual contaminant spread when the diffusion limit is attained, that is, $k\sigma \ll 1$ and $u\tau \ll 1$, so that the fractional diffusion equation is a good asymptotic expansion of CTRW. To assess this statement, we compare the analytical variance (\ref{variance_fde_sub}) with Monte Carlo simulation results for two values of $\alpha$. Monte Carlo estimates, which are obtained by simulating the microscopic dynamics of the particles, will be assumed as a reference solution of CTRW. In the first example, $10^5$ particles were followed up to a time $t_{\max}=1$, with the simulation parameters $\alpha=0.5$, $\sigma=0.5$, and $\tau=10^{-4}$ (see Fig.~\ref{fig_fde_sub1}). In this case the fractional diffusion prediction (\ref{variance_fde_sub}) agrees perfectly with the Monte Carlo simulation.

\begin{figure}[t]
\begin{center}
\scalebox{0.60}{\includegraphics{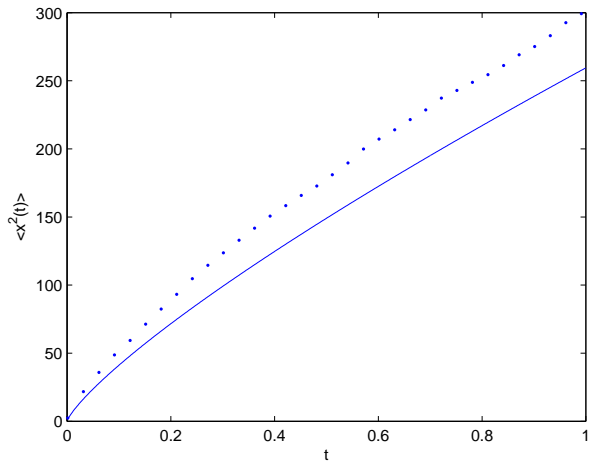}}
\caption{Comparison between the analytical (Eq.~(\ref{variance_fde_sub}), solid line) and Monte Carlo results (dots) as a function of time for $\alpha=0.8$. There is a clear disagreement.}
\label{fig_fde_sub2}
\end{center}
\end{figure}

Next we let $\alpha=0.8$, with $\sigma=0.5$, $\tau=10^{-4}$, and $t_{\max}=1$ (see Fig.~\ref{fig_fde_sub2}). In this case the fractional diffusion prediction (\ref{variance_fde_sub}) underestimates the actual Monte Carlo spread, so that fractional diffusion cannot be considered a good approximation of CTRW. The error introduced by the fractional diffusion prediction is far beyond the fluctuations due to the finite Monte Carlo statistics. We remark that in both cases the explored time scales are such that $\tau \ll t_{\max}$, thus ensuring that the diffusion limit has been attained.

What is the origin of the discrepancies shown in Fig.~\ref{fig_fde_sub2}? Straightforward but tedious calculations show that the expansion of $w(u)$ to second order in $u$ yields:
\begin{equation}
w(u)=1-c_\alpha(u\tau)^\alpha-c_1(u\tau)+ o (u^2),
\label{w_u_sub}
\end{equation}
where $c_1=\alpha/(\alpha-1)$. Therefore, when the exponent $\alpha$ is small, the linear term in $u$ in Eq.~(\ref{w_u_sub}) is expected to play no significant role in the limit $u\tau \ll 1$, thus justifying the accuracy of the fractional diffusion equation predictions. In contrast, if $\alpha \rightarrow 1$, the two terms in the expansion (\ref{w_u_sub}) become comparable and the effects of the linear contribution will no longer be negligible \cite{margolin, marseguerra_zoia1}. More precisely, when $\alpha<1$ the contribution of $c_1(u\tau)$ will always be subdominant with respect to $c_\alpha(u\tau)^\alpha$ (for $u \to 0$), so that after a sufficiently long time ($t \gg \tau$) we always recover the fractional diffusion equation as the correct asymptotic limit of CTRW. In this respect the term $c_1(u\tau)$ introduces pre-asymptotic corrections to the behavior of the subdiffusive particles. However, the values of $\alpha$ and $\tau$ are usually imposed by experiment and inherently linked to the choice of $w(t)$, that is, to the microscopic dynamics of the particles. It is therefore interesting to systematically explore the behavior of the CTRW and fractional diffusion solutions for intermediate time scales where the diffusion limit $u\tau \ll 1$ holds but the truncation of (\ref{w_u_sub}) to the first non-constant term might not be appropriate, depending on the value of $\alpha$. 

\begin{figure}[t]
\begin{center}
\scalebox{0.60}{\includegraphics{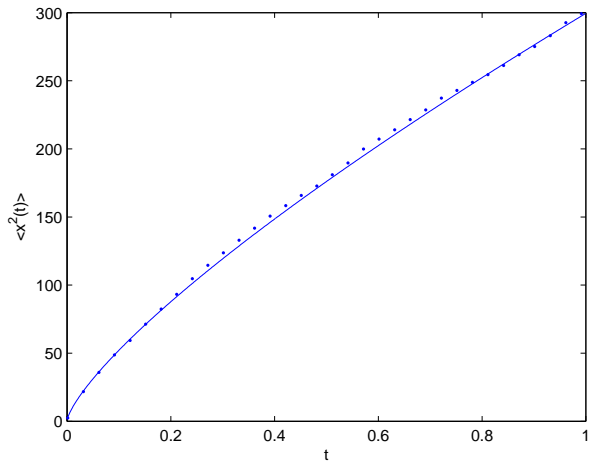}}
\caption{Comparison between the variance (\ref{FFPK_inv_moment}) as obtained by numerical inversion (solid line) and Monte Carlo simulations (dots) for $\alpha=0.8$. The two curves are in excellent agreement.}
\label{fig_fde_sub3}
\end{center}
\end{figure}

To quantitatively assess the relevance of the contribution $c_1(u\tau)$ in the expansion of $w(u)$, we substitute Eq.~(\ref{w_u_sub}) into the propagator (\ref{ctrw_pdf}), to obtain:
\begin{equation}
u P(k,u) -1=-\frac{\sigma^2 k^2}{c_\alpha \tau^\alpha u^{\alpha-1}+c_1\tau}P(k,u).
\label{propagator_fde_sub}
\end{equation}
By making use of the formal properties of the Riemann-Liouville operators, Eq.~(\ref{propagator_fde_sub}) may be finally recast into a modified fractional derivative equation
\begin{equation}
\frac{\partial}{\partial t} P(x,t) = D_\alpha \partial^{1-\alpha}_{0, t} \frac{\partial^2}{\partial x^2} P(x,t) + q_{\alpha,1} \partial^{1-\alpha}_{0, t} \left\lbrace \frac{\partial}{\partial t} P(x,t) \right\rbrace ,
\label{frac_transport}
\end{equation}
where $q_{\alpha,1}= c_1 \tau/c_\alpha \tau^\alpha$ \footnote{Note that the that the operators $\partial^{1-\alpha}_{0, t}$ and $\frac{\partial}{\partial t}$ do not commute, so that $\partial^{1-\alpha}_{0, t}\frac{\partial}{\partial t} \ne \partial^{2-\alpha}_{0, t}$.}. If $\alpha$ is sufficiently small, we can expand
\begin{equation}
\frac{1}{c_\alpha \tau^\alpha u^{\alpha-1}+c_1 \tau} =\frac{u^{1-\alpha}}{c_\alpha \tau^\alpha}\left( \frac{1}{1+q_{\alpha,1}u^{1-\alpha}}\right) \simeq \frac{u^{1-\alpha}}{c_\alpha \tau^\alpha} \left(1- q_{\alpha,1}u^{1-\alpha}\right) \simeq \frac{u^{1-\alpha}}{c_\alpha \tau^\alpha} + o(u^{2-2\alpha}),
\end{equation}
so that we recover the standard fractional diffusion equation. If instead $\alpha$ is of the order of unity, the two terms $c_\alpha(u\tau)^\alpha$ and $c_1(u\tau)$ (which are mirrored in the $D_\alpha \partial^{1-\alpha}_{0, t}$ and $q_{\alpha,1} \partial^{1-\alpha}_{0, t} \frac{\partial}{\partial t}$ operators, respectively) are in competition, and their specific contributions to the overall functional form of the propagator $P(x,t)$ cannot be neglected. We argue that Eq.~(\ref{frac_transport}), which is the central result of our work, also  provides a suitable description of the particle dynamics for pre-asymptotic time scales. To support this argument we compute the spread associated to Eq.~(\ref{frac_transport}). By definition,
\begin{equation}
\left\langle x^2(t) \right\rangle = {\cal L}^{-1} \left\lbrace \frac{2 \sigma^2}{u^2}\frac{1}{c_\alpha \tau^{\alpha} u^{\alpha-1} +c_1 \tau}\right\rbrace .
\label{FFPK_inv_moment}
\end{equation}
The inverse $\cal L$ transform appearing in Eq.~(\ref{FFPK_inv_moment}) can be computed numerically \cite{dehoog, hollenbeck}. The curves thus obtained are again compared with Monte Carlo simulation. Figure~\ref{fig_fde_sub3} clearly shows that the modified Eq.~(\ref{frac_transport}) provides a reliable estimate of the system evolution even at pre-asymptotic time scales.

\section{Conclusions}
\label{conclusions}

The fractional diffusion equation arises as an asymptotic approximation to an exact master equation and suffers from some limitations \cite{barkai}. As an example, although anomalous diffusion is often experimentally found to be a transient phase, which after a suitable time scale generally relaxes to Fickian diffusion, the fractional diffusion equation cannot take into account for this transition because the anomalous behavior is assumed to hold even at infinite time \cite{cortis_gallo, dentz_cortis, klafter1, klafter2, schneider}. Moreover, the CTRW formulation in terms of a generalized Fokker-Planck equation with a memory kernel recently proposed e.g. in \cite{cortis_gallo} can take into account more general expressions for $w(u)$ (as compared to to those allowed by the fractional derivative approach), as well as boundary and initial conditions in multiple dimensions. In this respect, the CTRW formalism, thanks to its greater flexibility, can capture and quantify more general tranport instances, as shown e.g. in \cite{cortis_ex1,cortis_ex2}.

It must be also emphasized that not all anomalous diffusion phenomena can be represented in a fractional derivative formalism, since anomalous diffusion is usually context-dependent many specific realizations do not fall within this framework (even though enlarged so to include pre-asymptotic corrections). This is the case, for instance, of the diffusion problems studied by \cite{cortis_knudby}, where slower than Fickian, but faster than algebraic decays are found, or by \cite{zoia_kardar}, where anomalous transport in a fracture network gives rise to a scaling which is at variance with the power-law predicted by the fractional derivative framework.

Nonetheless, a relevant feature of the fractional diffusion equation (in spite of its limitations) is that the fractional derivative formulation may easily include external fields in a simple manner and is naturally suitable for solving boundary value problems \cite{metzler_klafter, zoia}. In this respect, the fractional diffusion equation has been recently shown to act as a unifying framework for the quantitative description of different physical phenomena where anomalous diffusion plays a significant role \cite{klafter1, klafter2}. Moreover, many standard mathematical techniques for solving partial differential equations are readily applicable to the fractional diffusion equation.

In the context of particle transport in heterogeneous materials the fractional diffusion formulation may be used to analyze the evolution of the contaminant plume spread in time. Comparing the model prediction with measured experimental data could then lead to an estimate of the exponent $\alpha$ characterizing the system. It is therefore important to quantify the limit of validity of the fractional diffusion asymptotic subset. In this paper we have shown that, if $\alpha \to 1$, the pre-asymptotic corrections to fractional diffusion might significantly affect the model predictions. Their neglect would induce gross errors in the model which would distort the estimate of $\alpha$. To overcome this problem, we have derived a modified transport equation involving fractional derivatives: we have shown that the particle spread predicted by this model is in excellent agreement with Monte Carlo simulation also at the pre-asymptotic time scales. This modified equation might be suitable to explore the subdiffusive dynamics of physical systems close to the limit $\alpha \to 1$.

Finally, we remark that analogous pre-asymptotic corrections would come into play also in the case of diffusive dynamics ($\alpha>1$), if we were to adopt a power-law decaying distribution for the waiting times. These corrections would have a significant role close to the limit $\alpha \to 1$ and their origin is in the nature of the functional form of $w(t)$. Only, the role of the two contributions would be the opposite with respect to the case examined in this work: the dominant contribution would come from the term in $u$, the term in $u^\alpha$ being subdominant for $u \to 0$, as expected. As a particular case, it is also possible to choose the (arbitrary) waiting time distribution so that the coefficient $c_1$ is identically zero: a widely adopted choice is assuming a L\'evy stable law for $w(t)$, so that its Laplace transform expansion would not contain the term in $u$.

\begin{acknowledgments}

We wish to thank Prof. H. Gould for his precious comments and suggestions, which have substantially improved the quality of the present manuscript. A.Z. thanks the Fondazione Fratelli Rocca for financial support through a Progetto Rocca fellowship.

\end{acknowledgments}

\end{document}